# FACILE FABRICATION OF ASYMMETRIC SURFACES VIA MECHANOCHEMISTRY

Huihun Jung[1], Yusuf Nur[1,2], A. Kaan Kalkan[3], Melik C. Demirel[1]

[1]Materials Research Institute, and Department of Engineering Science and Mechanics, Pennsylvania State University, University Park, Pennsylvania 16802, USA

[2]Department of Chemistry, Mustafa Kemal University, 31070, Hatay, Turkey

[3]Functional Nanomaterials Laboratory, Oklahoma State University, Stillwater, Oklahoma 74078, USA

**ABSTRACT**

We demonstrate a mechanochemical approach to fabrication of a bidirectional surface on a doped silicon wafer. In the initial step of our fabrication procedure, a high-density polyethylene (HDPE) stylus is pressed and drawn on the Si surface. This rubbing action provides a mechanochemical treatment of the Si surface, which unexpectedly alters the etching direction in the subsequent metal-assisted chemical etching step. Hence, Si nanowires of two different orientations can be patterned conveniently by a rub-write step. The mechanism of bidirectional Si nanowire formation is investigated. This anisotropy is exploited to propel water droplets on a fluidic track.

Directional surfaces composed of asymmetric structures are widely used by Nature for wet and dry adhesion [1-4]. Inspired from Nature, several techniques were employed to synthesize asymmetric textured surfaces ranging from template and template-free deposition to photolithography to direct printing [5]. Engineered nanotextured surfaces were reported in transport studies [6] which derive their anisotropic adhesive wetting directly from their asymmetric nanoscale morphology. There is also a crucial need in microfluidics for scalable fabrication of directional surfaces with predefined geometrical and topographical features. Although lithographic techniques were used to fabricate directional textured surfaces [7,8], the number of reports on bi-directional wetting surfaces is very limited.

Here, we demonstrate fabrication and patterning of bidirectional surfaces on doped silicon wafers by wet etching. The present work employs the two-step metal-assisted chemical etching (MACE) for the preparation of Si nanowire (NW) arrays. MACE was initially demonstrated by Li and Bohn [9]. Because of its simplicity, reproducibility and yield of single crystalline NWs, MACE has received significant attention for potential applications. Here, we find it beneficial to review the mechanism of NW formation by MACE to better discuss how bidirectional NWs are obtained in the present work.

Following past MACE reports [10-16], we illustrate the reaction mechanisms (Reactions 1-4) in two-step MACE in Figure 1a. The instrumental reduction (1, 3 in blue), oxidation (2 in red) and chemical (4 in black) reactions are displayed. Corresponding electron conduction (i.e., through valence band states of p-type Si) from anodic to cathodic sites are also represented by arrows. Initial step of MACE involves silver ($Ag^+$) reduction in the form of nucleation and growth of Ag NPs [17,18]. Subsequently in the second step, these NPs catalyze etching of Si as well as serve as cathodic electrodes. The overall electromotive force for the redox couple in step 1 (Reactions 1 and 2) is 1.71 V (standard conditions) that explains its spontaneous occurrence. In the present demonstration, our Ag reduction has involved immersion of a p-type Si (111) or (100) wafer (Figures 1b and S1 respectively) into a 0.02 M $AgNO_3$ and 20% HF solution for 1 min. Subsequently, the sample is let to dry in the ambient air. In the second step, Reaction 3 replaces Reaction 1, where $H_2O_2$ replaces $Ag^+$ (i.e., $AgNO_3$) as the oxidant. Accordingly, the Ag-metallized Si has been immersed in $H_2O_2$:$H_2O$ (1:39.5) mixture with 20% HF for 30 min. The redox couple of Reactions 3 and 2 accounts for an electromotive force of 2.69 V (standard conditions). It is well known that etching of bare Si is limited in HF + $H_2O_2$ solutions [5]. Therefore, Reaction 3 requires the presence of Ag NPs as catalysts [13,14,16]. The most widely acknowledged model of MACE argues that, Si beneath the catalyst metal is etched at a faster rate than the Si without metal coverage. Consequently, the metal NP sinks into the Si, forming pores, and NWs upon coalescence of the pores (Figure 1a) [13]. It is also generally accepted that Si—Si bond cleavage occurs at the metal/Si interface by valence electron capture by the metal (i.e., hole injection to Si) [13]. These electrons are used by Reaction 3 at the metal/electrolyte interface. Subsequently, the freed Si atoms diffuse through the metal to the metal/electrolyte interface, where they are oxidized and the silicon oxide product is dissolved by HF [13]. The directional etching of the pores along <100> is a consequence of the crystal anisotropy and has been explained by back-bond breaking model [2,6]. NW formation is sporadically observed at the unpolished back side of the Si wafer, which we owe to its almost complete coverage by Ag (Fig S2). Upon sealing of the backside with epoxy, we observe a dramatic reduction in the etching rate at the polished surface. Hence, we infer the reduction (Reaction 3) dominantly occurs at the backside. This observation clearly suggests the hole injection to the Si at the etch front can be dominantly provided from the metallized backside of the wafer, where reduction is most efficient. We contemplate Reaction 3 also occurs at the NP/electrolyte interface in the pores, as generally conceived in the literature, but it is diffusion-limited due to constricted transport in the pores. Although the Si valence band is energetically well-aligned for Reaction 3 [13], reduction is understood to be limited on Si surfaces at the pores and NW tips due to a higher overpotential on Si [13] as well as reduced conductivity in the NWs [12].

SEM images (Fig 1c) acquired from four different quadrants of the Si wafer after MACE indicates the tilted NWs reproducibly point toward [010] indicative of etching occurs only in the direction of [0$\bar{1}$0] (given the cleaved surface at the flat of the wafer represents the (110) plane). On the other hand, upon mechanochemical treatment of the Si surface in the present work, the NWs were found to be aligned in the [001] direction. As depicted by Figure 2a, we performed this treatment on a narrow region of a p-type Si (111), immersed in dilute HF solution, prior to MACE (see supplementary material for experimental details). A high density polyethylene (HDPE) tweezer was pressed on the Si surface with an approximate pressure of 10 kPa and drawn (i.e., rubbed) along a straight line at a speed of about 10 cm/s. Figure 2b indicates the mechanochemically treated region does not appear different from its exterior, but the borders are noticeable. On the other hand, the treated and untreated regions become apparently different after the MACE. In Fig. 2c, the SEM image shows qualitatively that two regions differ in silver particle density (e.g., mechanochemical treated part has lower particle density) after the 1st step of MACE. Further, as aforementioned, a more striking difference is observed after the 2nd step of MACE, where the orientation of the NWs in the treated region are different, corresponding to the etching direction of $[00\bar{1}]$. The resulting bidirectional surface is clearly seen in Figure 2d, where the NWs are aligned along [001] in the treated region and [010] outside. Here, at the boundary between the treated and untreated areas the NWs coalesce due to different orientations. Therefore spiky bundles form which appear at a lighter tone due to higher electron emission in SEM. This dramatic difference can be seen in detail in Figs. 2e and 2f (outside and inside of the drawn area, respectively). Additionally, the direction of the NWs in the treated region is found to be independent of the drawing direction of the HDPE stylus on the Si wafer. As control studies, we showed that magnetic field (Fig S3), HF concentration (Fig S4) or etching steps (Fig S5) neither changes the direction of etching nor creates bidirectional surfaces.

It is established in the MACE literature that NWs on p-type Si (111) are etched at an angle to the Si surface [13] as also observed by us (e.g., Fig. 1b). However, NWs with controllably different orientations to the same surface (Fig. 2d) have not been observed earlier. This directional difference occurs due to a mechanochemical effect induced by the rubbing of HDPE probe against the Si surface. With the objective of elucidating this effect, we investigated the Si surface using XPS after HF treatment only and mechanochemical treatment in HF. Surface chemistry of crystalline Si after HF exposures is well documented in the literature. Once Si is immersed in dilute HF, native oxide layer on the surface is removed and Si—H bonds form. Additionally, Si—F and Si—$F_2$ bonds could also form, if the wafer is kept in the HF solution for sufficient length of time. Si—H is a metastable bond, and converts to Si—F or Si—$F_2$. Figure 3a shows the broader range XPS spectra of (111) Si surface with and without mechanochemical treatment. Figure 3b shows the F1s spectrum, which was resolved to two components at 685.2 (Si—F) and 686.2 (Si—$F_2$) eV. The Si—$F_2$ component becomes dominant and the Si—F is subdued after the mechanochemical treatment. No C peaks, indicative of HDPE residues, are detectable in the XPS spectrum. Additionally, Fig. 3c shows SEM-EDAX spectrum of the p-type Si (111) surface after MACE. In addition to mechanochemical treatment by rubbing in HF, we have explored alternative ways to change the surface chemistry of Si and thereby the etching direction. To this end, we conducted high concentration (10%) and long duration (1 h) HF exposures, but no change in NW alignment was observed after MACE (Fig S5). In a second experiment, the rubbing was performed using a graphite probe, but no NWs were observed in the treated region for this case (Fig S6).

Finally, the bidirectional surface prepared using mechanochemical treatment, was applied to directional transport of water droplets. The mechanochemically treated region, where Si NWs exhibit a different direction and appears as a darker line in Fig. 4a (Multimedia view), and it serves as a track to guide/propel a water droplet under vibrational excitation. For this demonstration, a sample was attached horizontally to a PASCO SF 9324 brand mechanical vibrator. The oscillations of the mechanical vibrator were kept constant at low amplitude (0.585 mm). The motion was recorded with a digital camcorder (Sony DCRTRV50), and the velocities were estimated from the movies using ImageJ and Windows Movie Maker. It is found that for a given droplet volume, there is a specific narrow frequency band in which the droplet could be propelled. The frequency required to move the DI-water droplets (18.2 MΩ cm) as a function of drop size is given in Fig. 4b. In particular, the droplet volume is found to be inversely proportional with the propelling frequency. Figure 4c shows the relation between the droplet velocity and normalized frequency, recorded for droplets of different volumes. The plots essentially indicate a similar relation independent of the droplet size, that is, a resonant behavior upon matching of the excitation and natural frequencies.

In summary, we demonstrated that mechanochemical treatment of the Si surface alters the etching direction during metal-assisted chemical etching. Mechanochemistry opens up a number of potentially fruitful avenues in the study of symmetry breaking. For example, crystal symmetry in Si ideally allows 3 different directions for the etch fronts to propagate below the surface of a (111) wafer: $[\bar{1}00]$, $[0\bar{1}0]$ and $[00\bar{1}]$. Interestingly, when MACE is performed, we observe the etching occurs persistently along $[0\bar{1}0]$ over the whole wafer (Fig. 1c). This observation was also reproducible for p-type (111) Si wafers, acquired from three different sources. On the other hand, we showed that the mechanochemical treatment changes the etching direction to $[00\bar{1}]$. Clearly, as elucidated by XPS (Fig. 3), the mechanochemical treatment also leads to a dramatic variation of the Si surface chemistry. Previous MACE studies have revealed a distance-dependent interaction between the Ag NPs whose origin is yet to be elucidated [13]. The NPs were observed to move collectively in a single etching direction, when they were sufficiently close, but the isolated NPs moved in multiple allowed directions (i.e., <100>) [13]. Although such interaction between NPs can explain their collective behavior, the selection of a single particular etching direction and its alteration by mechanochemical treatment must be linked to additional mechanisms. It was proposed in the literature [13], that the favorable etching direction also depends on the crystal morphology of the NPs (in addition to crystal structure of Si), because the catalytic activity depends on the crystal direction of the NP facets, which are in contact with the Si surface [13]. Based on similar reasoning, we anticipate the interface between the two crystals, Si and Ag, is of central importance to symmetry breaking in etching direction. We conjecture that when Ag NPs are reduced on Si in the first step of MACE, the metal adatom surface diffusion length may be comparable or larger than the NP size. As a result, the NPs can approach thermodynamic equilibrium and minimize their total surface energy resulting in a certain preferred crystal shape and orientation with respect to Si. Although Si allows three symmetric directions for etching, when interfaced with a particular Ag crystal plane and when catalytic etching is considered, this 3-fold symmetry is broken. Equivalently, the interface between the two crystals has increased anisotropy and decreased symmetry when MACE has to satisfy certain selection rules favored by both crystals. While the crystal orientation is not the same for all NPs, the etching occurs in a uniform direction, determined by the dominant crystal orientation of the Ag NPs due to the collective nature of etching, as discussed above. The preferred etching direction flips upon mechanochemical treatment, simply because the dominant crystal orientation of the Ag NPs change on the chemically modified Si surface (i.e., due to modified Ag/Si interface energies as shown in the XPS of Fig 3b). An indirect evidence for modified crystal orientation of the Ag NPs upon mechanochemical treatment is the change in particle density (Fig. 2c) but further experiments are required to elucidate this phenomenon, which will be the focus of our future research.

**SUPPLEMENTARY MATERIAL**

See supplementary material for details of experiments, movies and a set of supplementary SEM images as referred to in the paper.

**ACKNOWLEDGEMENTS**

This work was partially supported by the National Institutes of Health (grant # 1R21HL112114) and Pennsylvania State University. YN acknowledges financial support for this work from the Turkish Council of Higher Education.

**FIGURE CAPTIONS**

**FIGURE 1.** a) Schematic representation of the etching mechanism. The standard electrode potentials, E°, are with respect to standard hydrogen electrode. b) Cross-sectional SEM image of tilted NWs etched on a p-type Si (111) wafer. Scale bar is 5 μm. c) SEM images of etched surfaces sampled from four different quadrants of the p-type Si (111) wafer as indicated. Inset scale bars are 10 μm.

**FIGURE 2.** a) The mechanochemical treatment: HDPE stylus rubbed on the p-type Si (111) surface in HF solution. b) SEM image of the trace (in between dashed lines) formed after mechanochemical treatment. c) Comparison of Ag NPs reduced on the Si surface with and without mechanochemical treatment (in between dashed lines and outside, respectively). d) Bidirectional surface formed after step 2 of MACE. Higher resolution images of the NWs outside (e); and inside (f) of the mechanochemical treatment area.

**FIGURE 3.** XPS spectra of p-type Si (111) surface after HF treatment (top) and mechanochemical (MC) treatment (bottom) in HF (a and b). c) SEM-EDAX spectrum of the p-type Si (111) surface after MACE. The inset is the control (as received Si wafer) prior to MACE.

**FIGURE 4.** Application of the bidirectional surface. **a)** Bidirectional p-Si(111) wafer, where a water droplet is spotted on the region, whose textural orientation was controllably altered by mechanochemical treatment. Motion of a droplet on bidirectional surfaces is provided in the supplementary information. (Multimedia view) The Si surface looks green/yellow due to interference of light reflected from air/NW-layer and NW-layer/Si interfaces. **b)** Vibration frequency as a function of water droplet size for bidirectional surface demonstrates the inverse relationship between the frequency of vibration and water droplet size as water droplets are transported on the bidirectional track via vibrational frequencies at low amplitude. This is explained by Rayleigh's spherical volumes theory and can be quantified as $\omega=(3\pi\rho V/8\gamma)^{1/2}$, where V is the droplet volume, ω is the frequency of vibration, γ is water surface tension, and ρ is water density. **c)** Droplet transport speed on bidirectional surface as a functional of normalized vibration frequencies shows clustering of the data around the peak velocity. Normalized

frequency is given by $\omega/\omega^*$ or $\omega/(\gamma/m)^{1/2}$, where m is droplet mass, and $\omega^*$ is the natural frequency.

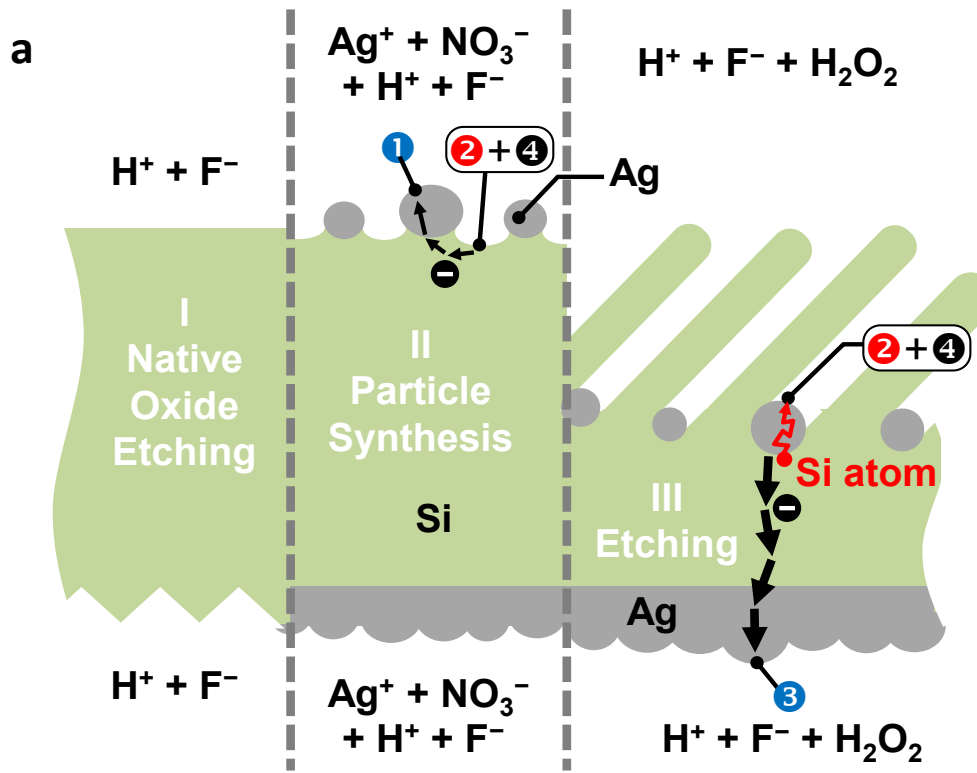
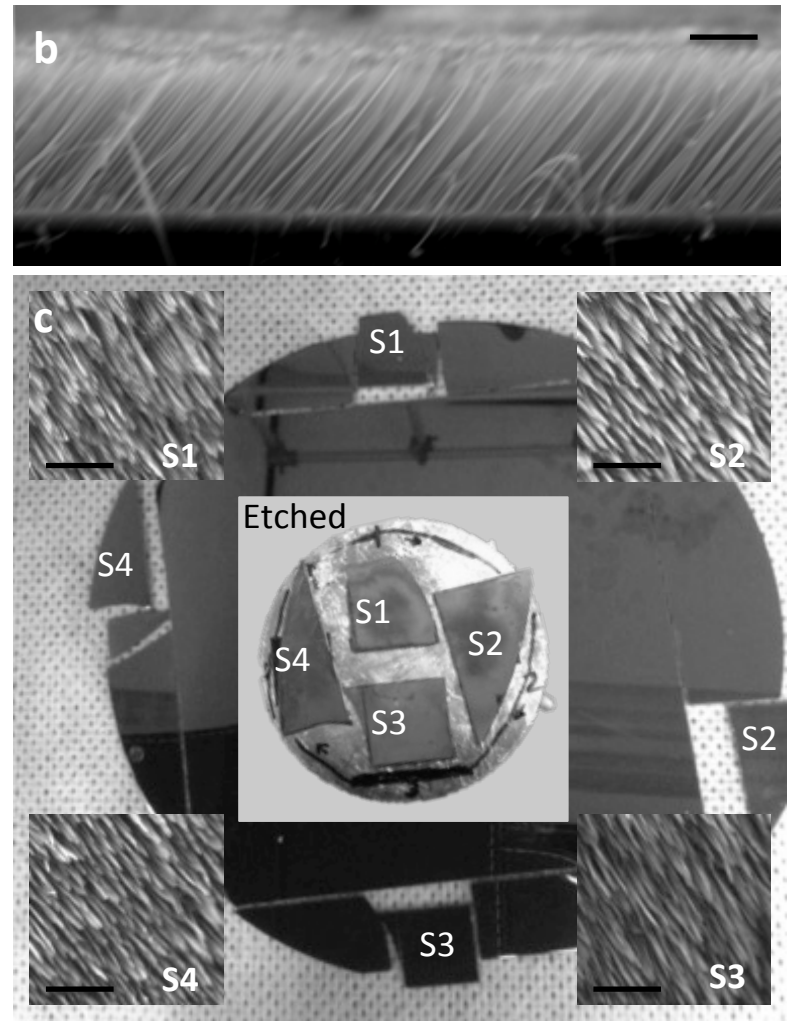

❶ $4Ag^+ + 4e^- \rightarrow 4Ag$  ($\Phi° = 0.80$ V)
❷ $Si + 2H_2O \rightarrow SiO_2 + 4H^+ + 4e^-$  ($\Phi° = 0.91$ V)
❸ $2H_2O_2 + 4H^+ + 4e^- \rightarrow 4H_2O$  ($\Phi° = 1.78$ V)
❹ $SiO_2 + 6HF \rightarrow [SiF_6]^{2-} + 2H_2O + 2H^+$

Figure-1

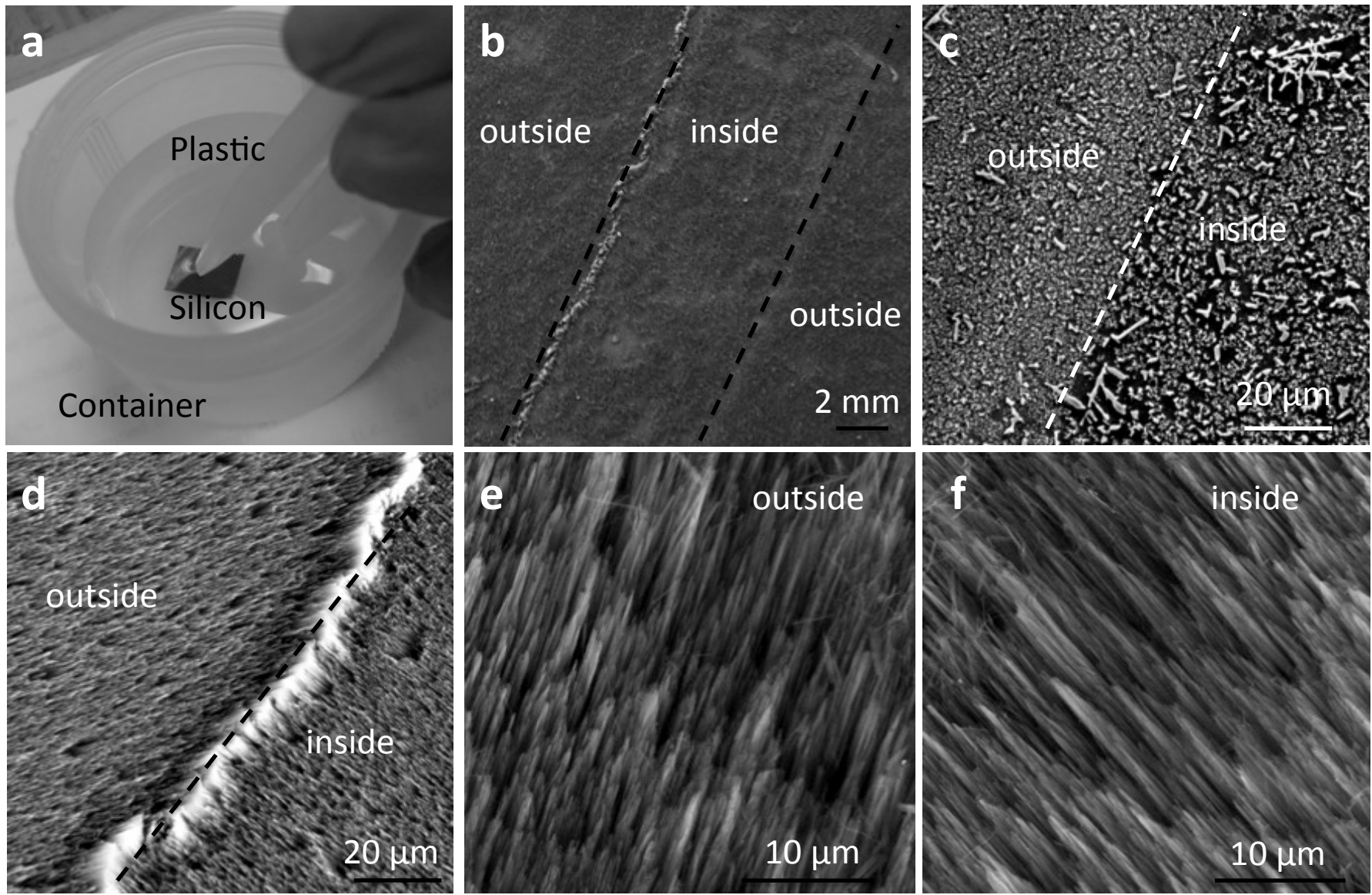

Figure-2

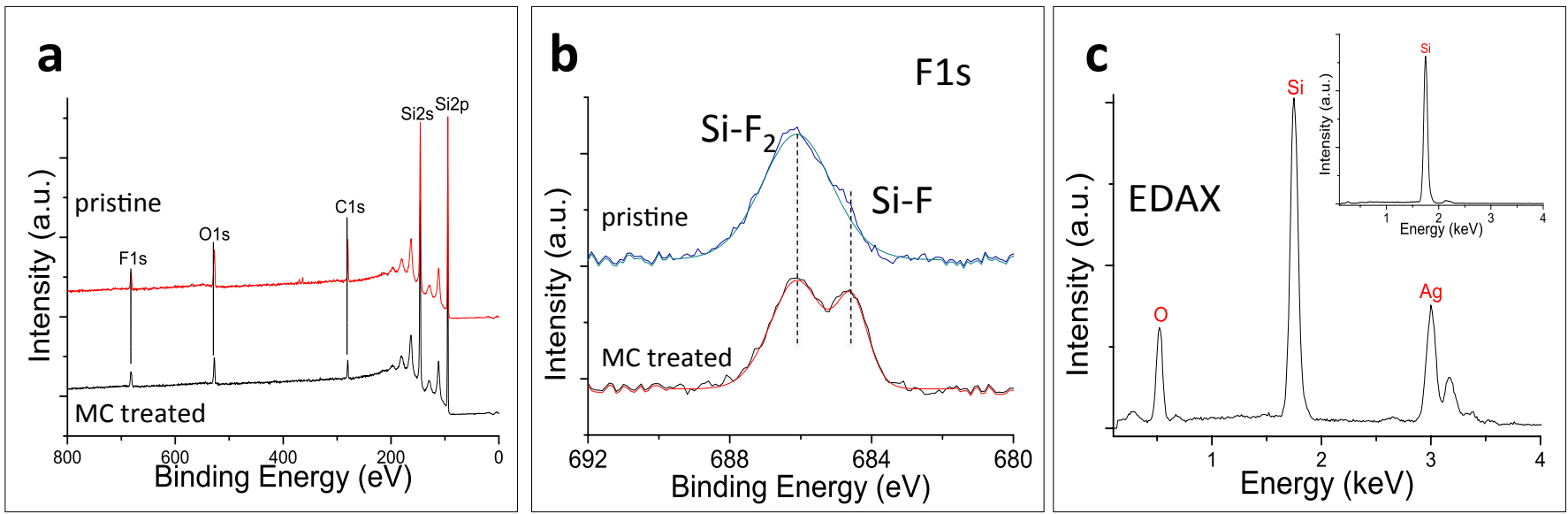

Figure-3

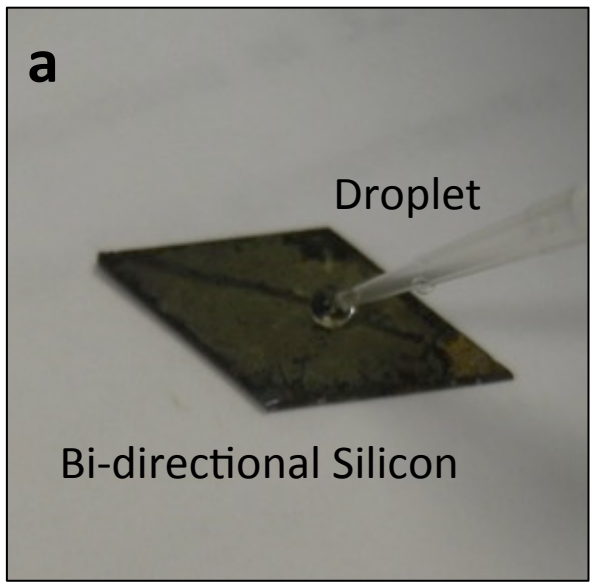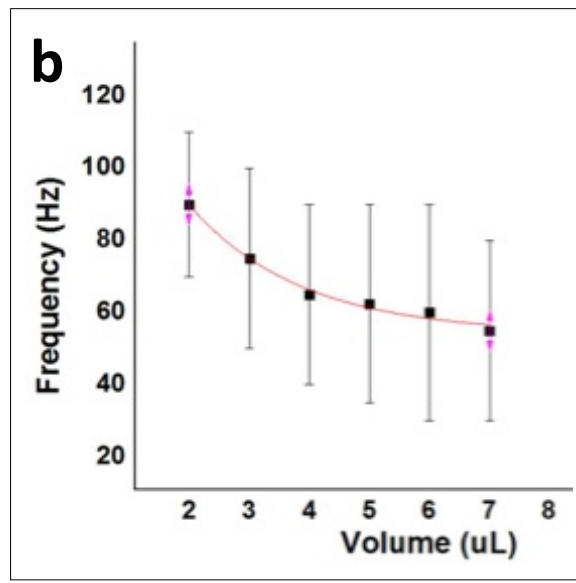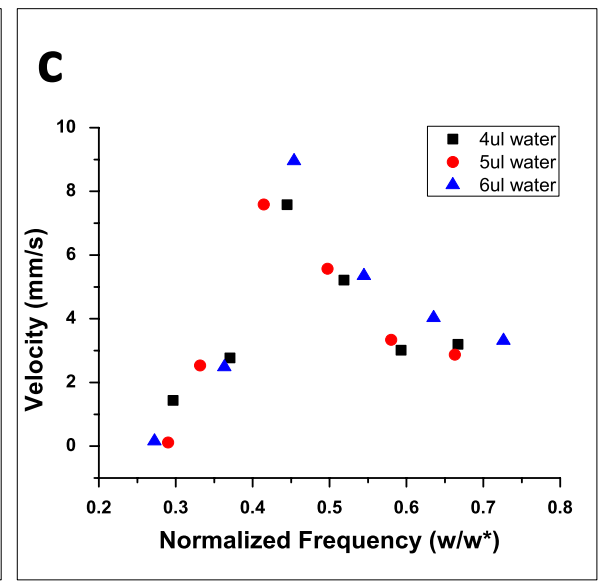